\documentclass[floatfix,showpacs,prb,aps,a4paper,twocolumn]{revtex4}

 \usepackage{bm}
 \usepackage{amsmath}
 \usepackage{amssymb}
 \usepackage{latexsym}
 \usepackage{amsfonts}
 \usepackage{epsfig}
 \usepackage{subfigure}
 \usepackage{color}
 \definecolor{darkblue}{rgb}{0,0,.5}
 \usepackage[linktocpage, colorlinks=true ,linkcolor=darkblue, citecolor=darkblue]{hyperref}
 \usepackage[all]{hypcap}
 \usepackage{upgreek}

 \newcommand{\expval}[1]{\left< #1 \right>}
 
 \newcommand{\ket}[1]{\left|#1\right>}

 \newcommand{\nn}{\nonumber\\}
 
 \newcommand{\f}[1]{\mbox{\boldmath$#1$}}

 \newcommand{\bea}{\begin{eqnarray}}
 \newcommand{\ea}{\end{eqnarray}}
 \newcommand{\eea}{\end{eqnarray}}
 
 \newcommand{\trace}[1]{{\rm Tr}\left\{ #1 \right\}}

 \newcommand{\abs}[1]{{\left| #1 \right|}}

 \newcommand{\HS}{\mathcal{H}_{\rm S}}
 \newcommand{\HB}{\mathcal{H}_{\rm B}}
 \newcommand{\HI}{\mathcal{H}_{\rm I}}
 \newcommand{\RS}{\rho_{\rm S}}

\newcommand{\ii}{\mathrm{i}}  
\newcommand{\ml}{\mu_{\rm L}}
\newcommand{\mr}{\mu_{\rm R}}
\newcommand{\mU}{\mu_{\rm U}}
\newcommand{\mD}{\mu_{\rm D}}
\newcommand{\gl}{\Gamma_{\rm L}}
\newcommand{\gr}{\Gamma_{\rm R}}
\newcommand{\gu}{\Gamma_{\rm U}}
\newcommand{\gd}{\Gamma_{\rm D}}

\newcommand{\bl}{\beta_{\rm L}}
\newcommand{\br}{\beta_{\rm R}}
\newcommand{\bu}{\beta_{\rm U}}
\newcommand{\bd}{\beta_{\rm D}}
\newcommand{\eB}{\upvarepsilon_{\rm B}}
\newcommand{\eA}{\upvarepsilon_{\rm A}}

\newcommand{\ddA}{d_{\rm A}^\dagger}
\newcommand{\dA}{d_{\rm A}}
\newcommand{\ddB}{d_{\rm B}^\dagger}
\newcommand{\dB}{d_{\rm B}}
\newcommand{\cU}{\chi_{\rm U1}}
\newcommand{\cD}{\chi_{\rm D1}}
\newcommand{\cL}{\chi_{\rm L1}}
\newcommand{\CL}{\chi_{\rm L2}}
\newcommand{\cR}{\chi_{\rm R1}}
\newcommand{\CR}{\chi_{\rm R2}}
\newcommand{\nU}{n_{\rm U1}}
\newcommand{\nD}{n_{\rm D1}}
\newcommand{\nL}{n_{\rm L1}}
\newcommand{\NL}{n_{\rm L2}}
\newcommand{\nR}{n_{\rm R1}}
\newcommand{\NR}{n_{\rm R2}}

\begin{document}

\title{Incomplete current fluctuation theorems for a four-terminal model}
\author{Thilo Krause}
\author{Gernot Schaller}\email{gernot.schaller@tu-berlin.de}
\author{Tobias Brandes}
\affiliation{Institut f\"ur Theoretische Physik, Technische Universit\"at Berlin, Hardenbergstr. 36, 10623 Berlin, Germany}

\begin{abstract}
We demonstrate the validity of the current fluctuation theorem for a double quantum dot surrounded by four terminals
within the Born-, Markov- and secular approximations beyond the Coulomb-blockade regime.
The electronic tunneling to two fermionic contacts conserves the total number of electrons, and the internal tunneling
is phonon-assisted by two bosonic baths.
Adapted choice of thermodynamic parameters between the baths may drive a current against an existing 
electric or thermal gradient.
We study the apparent violation of the fluctuation theorem when only some of the energy and matter currents are monitored.
\end{abstract}

\pacs{05.60.Gg,  
03.65.Yz 
}

\maketitle

Fluctuation theorems (FTs) connect forward and backward probabilities for processes associated
with a definite exchange of entropy~\cite{seifert2005a}.
Thereby they relate rather sophisticated and hard-to-calculate quantities with simple and universal thermodynamic
parameters, which constitutes part of their attractiveness.
When the matter and energy currents are tracked over a certain period of time, there exist 
simple versions of the current (or Full Counting Statistics) FT~\cite{andrieux2004a,andrieux2006a,andrieux2009a,esposito2007b}.
Significant progress made in the monitoring of electronic tunneling events through quantum dots (QDs)~\cite{gustavsson2006a,sukhorukov2007a} 
by using capacitively coupled quantum point contacts (QPCs) has led to an accurate understanding of Full Counting Statistics~\cite{flindt2009a}.
Unfortunately, the QPC signal originating from a monitored single dot does not allow to reconstruct bi-directional tunneling events.
Therefore, monitored double quantum dots allowing for bi-directional counting have entered the focus of interest~\cite{fujisawa2006a,utsumi2010a} 
and appear as ideal testbeds to check current FTs.

It is therefore essential to identify processes that may lead to modifications of the FT in an experimental setup.
The FT has been found to be modified due to true quantum effects such as Berry phases~\cite{ren2010a} or interference 
effects~\cite{saito2008a,foerster2008a,utsumi2009a}.
It may however also be modified due to detector back-action -- e.g., the back-action of QPC on a monitored double quantum dot~\cite{golubev2011a,kueng2011a} or
the influence of a single electron transistor monitoring another one~\cite{sanchez2010a,bulnes_cuetara2011a} -- or simply when detailed balance is explicitly broken via 
feedback control~\cite{schaller2011b}.

Here, we will argue at the example of phonon-assisted tunneling that an apparent violation of the current FT may also arise due to ignored couplings with
further baths (that might resist an experimental monitoring).

This paper is organized as follows:
In Sec.~\ref{sec:Model} we introduce our model and the method, followed by a verification of the multi-terminal
FT in Sec.~\ref{sec:FT}.
We discuss how the multi-terminal FT is modified when only partial information is available in Sec.~\ref{sec:incomplete}.


\section{Model and Methods}\label{sec:Model}


\subsection{Hamiltonian}

\begin{figure}[ht]
\begin{center}
\includegraphics[width=0.4\textwidth]{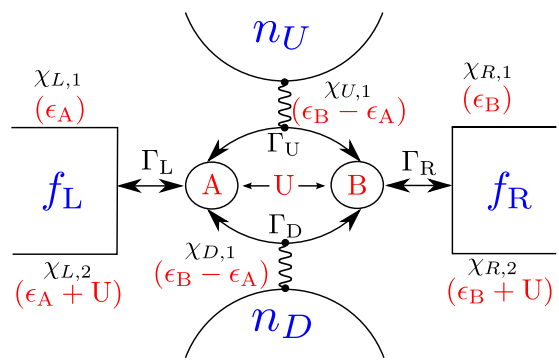}
\caption{\label{fig:Model}(Color online) 
Two quantum dots $A$ and $B$ are tunnel-coupled to adjacent fermionic contacts described by Fermi-distributions $f_L$ and $f_R$ 
with rates $\gl$ and $\gr$, respectively.
Electronic transitions between the QDs are only possible in the singly-charged sector via spontaneous emission or absorption of a boson with energy transfer $\eB-\eA$ 
from upper or lower phonon baths
(defined by Bose-Einstein distributions $n_U$ and $n_D$) with rates $\gu$ and $\gd$.
We explicitly allow for double occupation of the system, such that the coupling between particles and energy transferred to the fermionic reservoirs 
is not tight:
Depending on whether the inert QD is occupied or not, the energy exchanged (red, in brackets) with the fermionic baths may vary.
The use of multiple particle counting fields specific to both reservoir and transferred energy facilitates the calculation of the complete particle-energy counting
statistics.}
\end{center}
\end{figure}

We consider a double quantum dot system (see Fig.~\ref{fig:Model})
\bea
\label{eqn:systHamiltonian}
\HS=\eA \ddA \dA + \eB \ddB \dB + U \ddA \dA \ddB \dB
\eea
with on-site energies $\eA$ and $\eB$ and Coulomb-interaction $U$.
Without loss of generality (mirror symmetry) we assume that $\eA<\eB$.
The system is surrounded by two bosonic $\sigma\in\{U,D\}$ and two fermionic $\alpha\in\{L,R\}$ baths
\bea
 \HB=\sum_{k,\alpha}\upvarepsilon_{k,\alpha}c_{k,\alpha}^\dagger c_{k,\alpha}+\sum_{k,\sigma}\omega_{k,\sigma}b_{k,\sigma}^\dagger b_{k,\sigma}\,,
\eea
that are assumed to remain in thermal equilibrium throughout.
Each QD is coupled to its adjacent fermionic contact by the tunneling Hamiltonian
\bea\label{eqn:electronHamil}
\HI^{\rm el}=\dA\sum_{k}t_{k,\rm L}c_{k,\rm L}^\dagger+\dB\sum_{k}t_{k,\rm R}c_{k,\rm R}^\dagger+{\rm h.c.}\,,
\eea
where the tunneling amplitudes $t_{k,\alpha}$ lead to effective tunneling rates
$\Gamma_\alpha\equiv\Gamma_\alpha(\omega)=2\pi\sum_k \abs{t_{k,\alpha}}^2 \delta(\omega-\upvarepsilon_{k,\alpha})$ that
we assume to be energy independent (wideband limit).
In contrast, the transition  $A \leftrightarrow B$ is phonon-assisted via the interaction
\bea\label{eqn:phonIntHam}
 \HI^{\rm ph}=(\dA\ddB + \dB\ddA)\otimes\sum_{k,\sigma}(h_{k,\sigma}b_{k,\sigma}+h_{k,\sigma}^\ast b_{k,\sigma}^\dagger)\,,
\eea
such that -- under the rotating wave approximation -- an electron jump between the QDs goes with either emission or absorption of a boson from upper or lower bath~\cite{galperin2005a}.
We again summarize the corresponding amplitudes in effective energy-independent phonon-assisted electron tunneling rates 
$\Gamma_\sigma\equiv \Gamma_\sigma(\omega)=2\pi\sum_k \abs{h_{k,\sigma}}^2 \delta(\omega-\omega_{k,\sigma})$.
The resulting total model is described by the sum of all Hamiltonians ${\mathcal H}=\HS+\HB+\HI^{\rm el}+\HI^{\rm ph}$ and is visualized in 
Fig.~\ref{fig:Model}.


\subsection{Liouvillian}

We assume to be in the sequential tunneling regime, where second order perturbation theory in the couplings $t_{k\alpha}$ and $h_{k\sigma}$ to the contacts is a good approximation.
This regime can e.g.\ be achieved when all tunneling rates are small in comparison to the reservoir temperatures $\Gamma_{\alpha/\sigma} \ll k_{\rm B} T_{\alpha/\sigma}$~\cite{koenig1996a,emary2009a}.
More generally, Kondo physics is expected to be negligible when the reservoir temperatures are larger than the Kondo temperature~\cite{garate2011a}.
In this regime, performing the Born, Markov, and secular approximations~\cite{breuer2002} yields a master equation that is expected to yield valid results.
It involves only the system density matrix, is of Lindblad-form~\cite{lindblad1976a}, and in the (localized) system energy eigenbasis
($\HS\ket{00}=0$, $\HS\ket{10}=\eA\ket{10}$, $\HS\ket{01}=\eB\ket{01}$, and $\HS\ket{11}=(\eA+\eB+U)\ket{11}$) it assumes the form of a 
simple rate equation as long as the energy levels of $\HS$ are non-degenerate (recall that $\eB>\eA$).
In principle, the rates may be calculated rigorously but the result is also evident from Fermis Golden rule.
We assume the reservoirs to be in thermal equilibrium throughout, i.e., expectation values are given by the Fermi-Dirac
\bea\label{eqn:fermi}
\expval{c_{k,\alpha}^\dagger c_{k,\alpha}} \equiv 
f_\alpha(\varepsilon_{k,\alpha})=\frac{1}{e^{\beta_\alpha(\varepsilon_{k,\alpha}-\mu_\alpha)}+1}
\eea
or Bose-Einstein
\bea\label{eqn:bose}
\expval{b_{k,\sigma}^\dagger b_{k,\sigma}} \equiv
n_\sigma(\omega_{k,\sigma})=\frac{1}{e^{\beta_\sigma(\omega_{k,\sigma}-\mu_\sigma)}-1}
\eea
distributions, respectively, where $\beta_\alpha$ and $\beta_\sigma$ denote the inverse electronic or bosonic bath temperatures and
$\mu_\alpha$ and $\mu_\sigma$ the respective chemical potentials.
Writing the diagonal entries of the system density matrix in a vector $\RS=(\rho_{00,00}, \rho_{10,10},\rho_{01,01},\rho_{11,11})$, 
the master equation to this order assumes the form
$\dot{\RS} = \mathcal{L}\RS$, 
where the Liouvillian superoperators are given by
\bea\label{eqn:liouvillians}
\mathcal{L} &=& (\mathcal{L}_{\rm L}+\mathcal{L}_{\rm R}+\mathcal{L}_{\rm U}+\mathcal{L}_{\rm D})\,,\nn
\mathcal{L}_{\rm L}&=&\gl
\begin{pmatrix} 
-f_{\rm L} & +(1-f_{\rm L}) & 0 & 0 \\
+f_{\rm L} & -(1-f_{\rm L}) & 0 & 0\\
0 & 0 & -\bar{f}_{\rm L} & +(1-\bar{f}_{\rm L})\\
0 & 0 & +\bar{f}_{\rm L} & -(1-\bar{f}_{\rm L})
\end{pmatrix}\,,\nn
\mathcal{L}_{\rm R}&=&\gr
\begin{pmatrix} 
-f_{\rm R} & 0 & +(1-f_{\rm R}) & 0\\
0 & -\bar{f}_{\rm R} & 0 & +(1-\bar{f}_{\rm R})\\
+f_{\rm R} & 0 & -(1-f_{\rm R}) & 0\\
0 & +\bar{f}_{\rm R} & 0 & -(1-\bar{f}_{\rm R}
\end{pmatrix}\,,\nn
\mathcal{L}_{\rm \sigma}&=&\Gamma_{\sigma}
\begin{pmatrix} 
0 & 0 & 0 & 0\\
0 & -n_{\sigma} & +(1+n_{\sigma})& 0\\
0 & +n_{\sigma} & -(1+n_{\sigma})& 0\\
0 & 0 & 0 & 0
\end{pmatrix}\,.
\eea
Here, we have used the abbreviations
\bea
n_\sigma  &\equiv&  n_\sigma(\eB-\eA)\,,\nn
f_{\rm L} &\equiv&  f_{\rm L}(\eA)\,,\qquad f_{\rm R} \equiv f_{\rm R}(\eB)\,,\nn
\bar{f}_{\rm L} &\equiv&  f_{\rm L}(\eA+U)\,,\qquad \bar{f}_{\rm R} \equiv f_{\rm R}(\eB+U)\,,
\eea
compare Eqns.~(\ref{eqn:fermi}) and~(\ref{eqn:bose}).
In the Coulomb-blockade limit, the doubly occupied state always decays, which is formally expressed by the limits
$\bar{f}_L \to 0$ and $\bar{f}_R \to 0$, and the top-left $3\times 3$ submatrix of the above equation reproduces
previous models in the literature~\cite{rutten2009a}.

It should be noted that the above Liouvillian~(\ref{eqn:liouvillians}) only satisfies local detailed balance~\cite{esposito2010a}:
When the system is only coupled to a single junction $\Sigma\in\{L,R,U,D\}$, detailed balance 
$\mathcal{L}^{k\ell}_\Sigma \bar\rho^\ell_\Sigma = \mathcal{L}^{\ell k}_\Sigma \bar\rho^k_\Sigma$ -- where $\bar\rho^k_\Sigma$ denotes the corresponding
single-junction stationary state -- is obeyed.
When it is coupled to all four terminals however, global detailed balance is broken.


\subsection{Conditional Master Equation}

Being responsible for transitions between different system states, the off-diagonal matrix elements of the Liouvillians
in Eqns.~(\ref{eqn:liouvillians}) can be used to set up a connected set of equations for density matrices $\rho^{(n)}(t)$ conditioned upon the
number of particles $n$ that have tunneled to a respective reservoir~\cite{gurvitz1996a}.
That is, although the state of the system density matrix does neither yield information on the number of particles nor the 
energy transferred to a certain bath, the full time-record of a single trajectory would reveal this information.
When we only focus on a single reservoir to and out of which we count all particle jumps, the conditioned density matrices follow an equation of the 
form 
$\dot{\rho}^{(n)} = {\mathcal L}_0 \rho^{(n)} + {\mathcal L}_+ \rho^{(n-1)} + {\mathcal L}_- \rho^{(n+1)}$, 
where ${\mathcal L}_+$ (${\mathcal L}_-$) describes particle jumps into (out of) the reservoir of interest and 
${\mathcal L}_0$ contains the remaining terms (jumps to and from other reservoirs as well as the diagonal matrix elements).
Alternatively, such $n$-resolved master equations may also be obtained by performing the derivation after a detector has been added to the system~\cite{schaller2009b}.
By performing a Fourier transform
\bea\label{eqn:Fourier}
\RS(\chi,t)\equiv\sum_{n}\RS^{(n)}(t)e^{\ii n \chi}\,,
\eea
one can convert the infinitely large $n$-resolved, conditional master equation to a four-dimensional one at the price of introducing the
counting field $\chi$.
Being e.g.\ interested in the number of particles in the left contact, this formally corresponds to the replacements $f_L\to f_L e^{-i\chi}$
and $(1-f_L) \to (1-f_L) e^{+i\chi}$ in the off-diagonal matrix elements of Eqns.~(\ref{eqn:liouvillians}), respectively.

However, here we are interested in the full energy-particle counting statistics, such that one has to treat e.g.\ single-electron jumps corresponding
to an energy change of $\eA$ and to an energy change of $\eA+U$ in the system differently.
Therefore, we introduce two counting fields for each fermionic contact and one for each bosonic contact.
Note that in the Coulomb-blockade limit, energy and particle fluxes are tightly coupled (often also called strongly coupled~\cite{rutten2009a}), such that also for the
fermionic contacts a single counting field would suffice.
The Liouvillian as a function of energy-resolved counting fields reads
\bea\label{eqn:FCSLiouvillian}
\mathcal{L}(\f{\chi}) &\equiv& \mathcal{L}(\cL,\CL,\cR,\CR,\cU,\cD)\nn
&=& \mathcal{L}_{\rm L}(\cL,\CL)+\mathcal{L}_{\rm R}(\cR,\CR)\nn
&&+\mathcal{L}_{\rm U}(\cU)+\mathcal{L}_{\rm D}(\cD)\,,
\eea
where the explicit counting field dependence can be obtained from Eqns.~(\ref{eqn:liouvillians}) by performing the replacements
\bea
f_{\rm L} &\to& f_{\rm L} e^{-\ii\cL}\,,\qquad (1-f_{\rm L}) \to (1-f_{\rm L}) e^{+\ii \cL}\,,\nn
\bar{f}_{\rm L} &\to& \bar{f}_{\rm L} e^{-\ii\CL}\,,\qquad (1-\bar{f}_{\rm L}) \to (1-\bar{f}_{\rm L}) e^{+\ii \CL}\,,\nn
f_{\rm R} &\to& f_{\rm R} e^{-\ii\cR}\,,\qquad (1-f_{\rm R}) \to (1-f_{\rm R}) e^{+\ii \cR}\,,\nn
\bar{f}_{\rm R} &\to& \bar{f}_{\rm R} e^{-\ii\CR}\,,\qquad (1-\bar{f}_{\rm R}) \to (1-\bar{f}_{\rm R}) e^{+\ii \CR}\,,\nn
n_{\rm U} &\to& n_U e^{-\ii \cU}\,,\qquad (1+n_{\rm U}) \to (1+n_{\rm U}) e^{+\ii \cU}\,,\nn
n_{\rm D} &\to& n_D e^{-\ii \cD}\,,\qquad (1+n_{\rm D}) \to (1+n_{\rm D}) e^{+\ii \cD}
\eea
in the off-diagonal matrix elements of the Liouvillian.
The counting fields are thereby related to specific transition frequencies in the system
\bea
\cL&\leftrightarrow&\eA\,,\qquad
\CL\leftrightarrow\eA+\rm U\,,\nonumber\\
\cR&\leftrightarrow&\eB\,,\qquad
\CR\leftrightarrow\eB+\rm U\,,\nonumber\\
\cU&\rightarrow&\eB-\eA\,,\nonumber\\
\cD&\rightarrow&\eB-\eA\,.
\eea
Therefore, we can now check the FT for fermions and bosons not only separately~\cite{harbola2007a}, but in a single model where
both species interact.


\subsection{Full Counting Statistics}

From the Liouvillian~(\ref{eqn:FCSLiouvillian}), one obtains the full statistics in a specific reservoir-energy channel by
taking derivatives with respect to the counting field, e.g.\ for the moments
\bea
 \expval{n_i^{k}(t)}\,=(-\ii \partial_{\chi_i})^{k}\left.\mathcal{M}(\f{\chi},t)\right|_{\f{\chi}=\f{0}}\,,
\eea
with the Moment-Generating-Function (MGF)
\bea\label{eqn:MGF}
 \mathcal{M}(\f{\chi},t)=\trace{e^{\mathcal{L}(\f{\chi}) t}\bar \rho}\,,
\eea
where $\bar \rho$ is the stationary density matrix defined by ${\mathcal L}(\f{0})\bar\rho=\f{0}$.

Most simple, the current as the time derivative of the first moment can be evaluated as (see e.g.~\cite{flindt2005a})
\bea\label{eqn:current}
\expval{\f{I}} = -\ii \trace{\left.\f{\partial_\chi} \mathcal{L}(\f{\chi})\right|_{\f{\chi}=\f{0}} \bar\rho}\,.
\eea
It is generally much more difficult to reconstruct the full probability distribution from the MGF as this involves the inverse
Fourier transform of the MGF -- compare Eq.~(\ref{eqn:Fourier})
\bea\label{eqn:probability}
 P_{\f{n}}(t)=\frac{1}{(2 \pi)^6}\int\limits_{-\pi}^{\pi}{\rm d}^{6}\f{\chi}\mathcal{M}(\f{\chi},t)e^{-\ii \f{n} \cdot  \f{\chi}}\,.
\eea
Provided a situation where $\lambda_1(\f{\chi})$ is the only eigenvalue of the Liouvillian with $\lambda_1(\f{0})=0$ (but
see e.g.~\cite{schaller2010b} for a more generalized treatment)
the cumulant-generating function (CGF) becomes linear in time~\cite{sukhorukov2007a}
\bea\label{eqn:CGF}
\mathcal{C}(\f{\chi},t) \equiv \ln\left[\mathcal{M}(\f{\chi}, t)\right] \to \lambda_1(\f{\chi}) t\,,
\eea
in the long-term limit (which will be denoted by an arrow further-on).
Therefore, $\lambda_1(\f{\chi})$ can be interpreted as the long-term CGF for the current.


\section{Complete Fluctuation Theorem}\label{sec:FT}

The characteristic polynomial of the Liouvillian~(\ref{eqn:FCSLiouvillian}) is given by
$\mathcal{D}(\lambda,\f{\chi})\equiv\det[\mathcal{L}(\f{\chi})-\lambda \f{1}]$ (not shown for brevity).
We have found that it fulfills the analytic property
\bea
\mathcal{D}(\lambda,-\f{\chi})=\mathcal{D}(\lambda,\f{\chi}+\ii \f{\Delta})\,,
\eea
where for all $\lambda$ the shift is given by
\bea\label{eqn:shift}
\f{\Delta} =\left(
\begin{array}{c}
\bl(\eA-\ml)\\
\bl(\eA+{\rm U}-\ml)\\
\br(\eB-\mr)\\
\br(\eB+{\rm U}-\mr)\\
\bu(\eB-\eA-\mU)\\
\bd(\eB-\eA-\mD)
\end{array}
\right)\,.
\eea
The characteristic polynomial can be decomposed as
$\mathcal{D}(\lambda,\f{\chi})=\prod\limits_{i=1}^4 \left[\lambda-\lambda_i(\f{\chi})\right]$ for all $\lambda$, where
$\lambda_i(\f{\chi})$ denote the four eigenvalues of Liouvillian~(\ref{eqn:FCSLiouvillian}).
It therefore
follows that all eigenvalues must obey the same symmetry.
In particular, we have also for the dominant eigenvalue -- the CGF for the current -- the symmetry relation
\bea\label{eqn:symmetryI}
\lambda_1(-\f{\chi})=\lambda_1(\f{\chi}+\ii \f{\Delta})\,,
\eea
with the same shift as in Eq.~(\ref{eqn:shift}).

The current FT is given by the ratio of forward and backward probabilities for particle and energy exchange with multiple baths in the long-term limit.
For the 4-terminal model considered here (see Fig.~\ref{fig:Model}), it follows from symmetry~(\ref{eqn:symmetryI}) by basic properties
of the inverse Fourier transform (see, e.g.\ the appendixes of Ref.~\cite{esposito2009a} for a more detailed discussion) 
and reads
\bea\label{eqn:FullFT}
\lim_{t \to \infty}\frac{P_{+\f{n}}(t)}{P_{-\f{n}}(t)}&=&
e^{\nL\bl(\eA-\ml)} e^{\NL\bl(\eA+\rm U-\ml)}\times\nn
&&\times e^{\nR\br(\eB-\mr)} e^{\NR\br(\eB+\rm U-\mr)}\times\nn
&&\times e^{\nU\bu(\eB-\eA-\mU)}\times\nn
&&\times e^{\nD\bd(\eB-\eA-\mD)}\,.
\eea
Using that transferred energies and particle numbers are related by
\bea
n_L &=& \nL+\NL\,,\qquad
n_R = \nR+\NR\,,\nn
n_U &=& \nU\,,\qquad
n_D = \nD\,,\nn
\Delta E_L &=& \nL \eA + \NL (\eA+U)\,,\nn
\Delta E_R &=& \nR \eB + \NR (\eB+U)\,,\\
\Delta E_U &=& \nU (\eB-\eA)\,,\qquad
\Delta E_D = \nD (\eB-\eA)\,,\nonumber
\eea
we find that the result~(\ref{eqn:FullFT}) is completely consistent with predictions in the literature~\cite{campisi2011a}
as one would expect for an effective rate equation satisfying local detailed balance.


\section{Incomplete Fluctuation Theorems}\label{sec:incomplete}

It is evident that the numbers of particles counted at all junctions are not independent.
The total number of electrons is conserved for example.
This results in further analytic properties of the characteristic polynomial ${\cal D}(\lambda,\chi)$, which transfer to
the long-term CGF~(\ref{eqn:CGF}).
We note here the identity
\bea\label{eqn:symmetryII}
\lambda_1(\cL,\CL,\cR,\CR,\cU,\cD) &=& \lambda_1(\cL+s_1-s_{\rm L},\nn
&&\;\;\;\CL+s_2-s_{\rm L},\nn
&&\;\;\;\cR+s_1+s_{\rm R},\nn
&&\;\;\;\CR+s_2+s_{\rm R},\nn
&&\;\;\;\cU+s_{\rm L}+s_{\rm R},\nn
&&\;\;\;\cD+s_{\rm L}+s_{\rm R})\nn
\eea
for arbitrary shifts $s_1$, $s_2$, $s_{\rm L}$, and $s_{\rm R}$.

It is also obvious that the structure of the two bosonic Liouvillians is identical and that their dependence on the respective 
bath occupation is linear. 
This implies that once one is only interested in e.g.\ the total number of bosons, the impact of the two reservoirs adds up to
a hypothetical single reservoir at some average occupation, similar to previous findings~\cite{schaller2011a}.
This leads to an additional analytic property of the characteristic polynomial, which also transfers to the dominant 
eigenvalue
\bea\label{eqn:symmetryIII}
\lambda_1(\cL,\CL,\cR,\CR,\chi_{\rm ph} + \ii \bu(\eB-\eA-\mU),&&\nn
\chi_{\rm ph} +\ii \bd(\eB-\eA-\mD)) &=&\nn
\lambda_1\Big(\cL,\CL,\cR,\CR,&&\nn
\chi_{\rm ph}+\ii \ln\left[\frac{1+\bar{n}}{\bar{n}}\right], \chi_{\rm ph}+\ii \ln\left[\frac{1+\bar{n}}{\bar{n}}\right]\Big)&&\,,
\eea
where the average bosonic occupation is simply given by the weighted sum
\bea
\bar{n} = \frac{\gu n_{\rm U}(\eB-\eA) + \gd n_{\rm D}(\eB-\eA)}{\gu+\gd}\,,
\eea
which allows one to define an average boson temperature at vanishing boson chemical potentials via 
\mbox{$\bar\beta_{\rm ph} (\eB-\eA) = \ln\left[\frac{1+\bar{n}}{\bar{n}}\right]$}.


\subsection{Electronic Transfer FT}

We are interested in the joint probability that $n_{\rm el}$ electrons have left the system at the right junction
(i.e., $\nR+\NR=+n_{\rm el}$) and $n_{\rm el}$ electrons have entered the system at the right junction (i.e., $\nL+\NL=-n_{\rm el}$).
This probability can in the long-time limit be evaluated via
\bea
P_{n_{\rm el}}(t) &=& \sum_{\stackrel{\nR,\NR}{\nR+\NR=n_{\rm el}}} \sum_{\stackrel{\nL,\NL}{\nL+\NL=-n_{\rm el}}} \left[\sum_{\nU,\nD} P_{\f{n}}(t)\right]\nn
&\to& \frac{1}{4\pi^2} \int\limits_{-\pi}^{+\pi} 
e^{\lambda_1(\chi_L,\chi_L,\chi_R,\chi_R,0,0) t -\ii n_{\rm el} (\chi_R-\chi_L)} d^2\chi\,,\nn
\eea
where we have eliminated the integrals by using that $\sum_n e^{\ii n \chi} = 2\pi \delta(\chi)$.
In order to relate the backward probability to the above equation, we consider in the limit where the 
electron temperatures are equal $\beta_L=\beta_R\equiv\beta_{\rm el}$
the identities 
\bea
\lambda_1(-\chi_L,-\chi_L,-\chi_R,-\chi_R,0,0) &&\stackrel{{\rm Eq.}~(\ref{eqn:symmetryI})}{=}\nn
\lambda_1\Big(\chi_L+\ii \beta_{\rm el}(\eA-\ml),\chi_L+\ii\beta_{\rm el}(\eA+U-\ml),&&\nn
\chi_R+\ii\beta_{\rm el}(\eB-\mr) ,\chi_R+\ii\beta_{\rm el}(\eB+U-\mr),&&\nn
\ii\bu(\eB-\eA-\mU),\ii\bd(\eB-\eA-\mD)\Big)&&\stackrel{{\rm Eq.}~(\ref{eqn:symmetryIII})}{=}\nn
\lambda_1\Big(\chi_L+\ii \beta_{\rm el}(\eA-\ml),\chi_L+\ii\beta_{\rm el}(\eA+U-\ml),&&\nn
\chi_R+\ii\beta_{\rm el}(\eB-\mr) ,\chi_R+\ii\beta_{\rm el}(\eB+U-\mr),&&\nn
\ii\bar\beta_{\rm ph}(\eB-\eA),\ii\bar\beta_{\rm ph}(\eB-\eA)\Big)&&\stackrel{{\rm Eq.}~(\ref{eqn:symmetryII})}{=}\nn
\lambda_1\Big(\chi_L+\ii \beta_{\rm el}(\eA-\ml+U/2)+\frac{\ii}{2} \bar\beta_{\rm ph}(\eB-\eA),&&\nn
\chi_L+\ii \beta_{\rm el}(\eA-\ml+U/2)+\frac{\ii}{2} \bar\beta_{\rm ph}(\eB-\eA),&&\nn
\chi_R+\ii \beta_{\rm el}(\eB-\mr+U/2)-\frac{\ii}{2} \bar\beta_{\rm ph}(\eB-\eA),&&\nn
\chi_R+\ii \beta_{\rm el}(\eB-\mr+U/2)-\frac{\ii}{2} \bar\beta_{\rm ph}(\eB-\eA),&&\nn
0,0\Big)&&.
\eea
Performing a shift of the integration variables in the denominator, we obtain the electronic transfer FT
\bea\label{eqn:electronic_transfer}
\frac{P_{+n_{\rm el}}(t)}{P_{-n_{\rm el}}(t)} &\to& e^{n_{\rm el}\left[\beta_{\rm el} (\eB-\eA+V)-\bar\beta_{\rm ph}(\eB-\eA)\right]}\,,
\eea
where $V\equiv\ml-\mr$ denotes the conventional bias voltage.
That is, with incomplete information, the full fluctuation theorem acquires the form of a modified FT with a shift term in the exponential.
Note also that the electronic transfer fluctuation theorem does not depend on the Coulomb interaction term $U$.

This has to be contrasted with modifications of the FT in the literature where one only observes a modified temperature~\cite{utsumi2010a,golubev2011a,kueng2011a}.
Instead, the apparent violation of the FT effectively mimics one found for a Maxwell demon model~\cite{schaller2011b}.

The same FT as in Eq.~(\ref{eqn:electronic_transfer}) would be obtained if e.g., only the number of electrons leaving at the right junction
was counted.
This is a consequence of electron number conservation, which is formally expressed by the symmetry 
\mbox{$\lambda_1(\chi_L,\chi_L,\chi_R,\chi_R,0,0)=\lambda_1(0,0,\chi_R-\chi_L,\chi_R-\chi_L,0,0)$} -- see also Eq.~(\ref{eqn:symmetryII}).

\begin{figure}[ht]
\includegraphics[width=0.48\textwidth,clip=true]{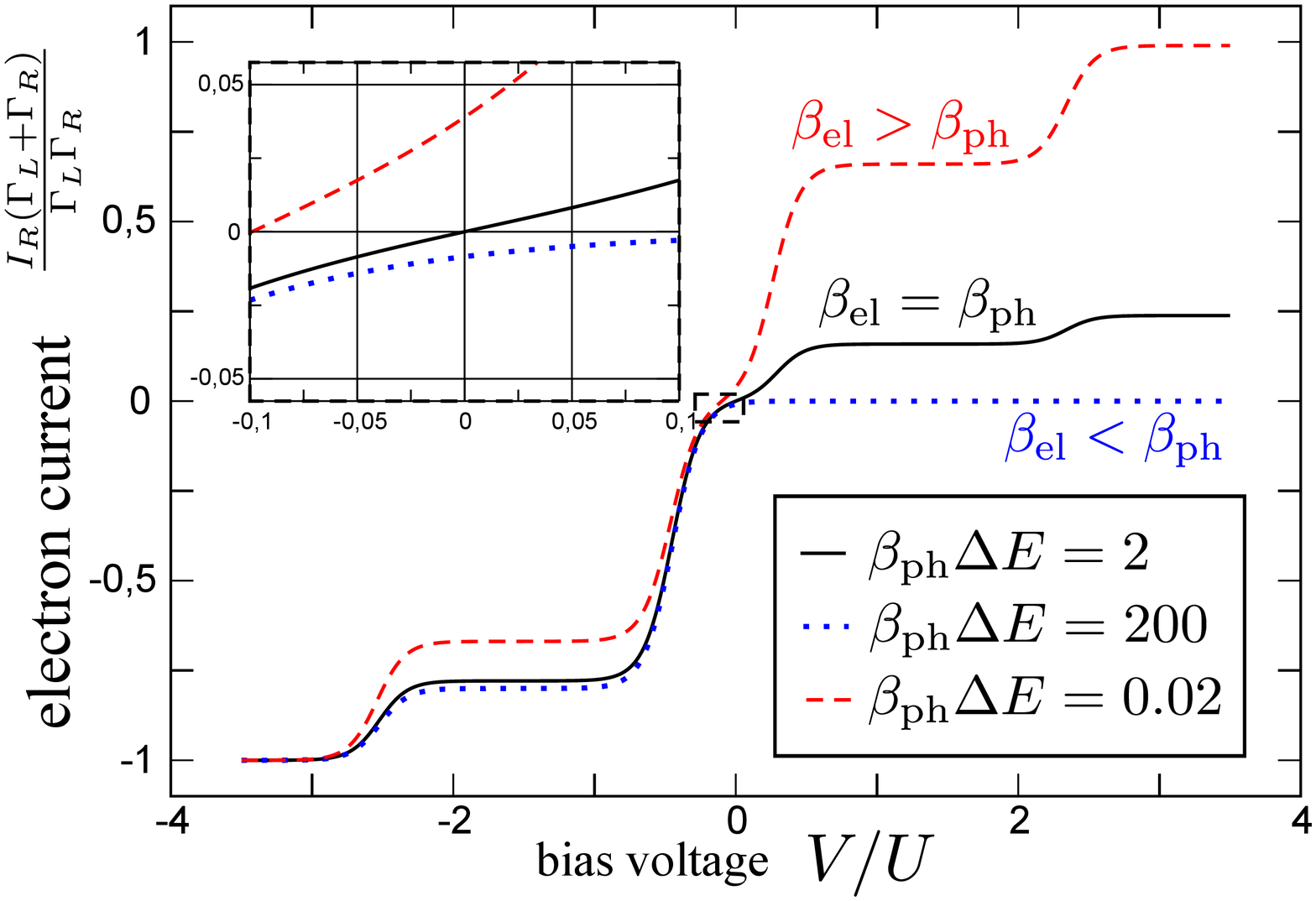}
\caption{\label{fig:Phon2in1}
(Color Online)
Renormalized electron current $I_R=I_{R1}+I_{R2}$ from Eq.~(\ref{eqn:current}) versus bias voltage for different phonon bath temperatures.
When the boson temperature exceeds the electron temperature, the current at zero bias voltage is positive (dashed red curve), whereas the opposite is
true for larger boson than electron temperatures (dotted blue curve) -- compare the zoomed inset.
Parameters have been chosen as $\beta_{\rm el}\Delta E \equiv \beta_{\rm el} (\eB-\eA)= 2$, $\beta_{\rm el} U=20$, $\mu_L=+V/2=-\mu_R$, $\mu_U=\mu_D=0$, $\gl=\gr=\gu=\gd$.
}
\end{figure}

The shift term in the FT has the interesting consequence that e.g.\ at zero bias voltage $V=0$, a current may still be generated from
left to right ($\beta_{\rm el}>\beta_{\rm ph}$) or vice versa ($\beta_{\rm el}<\beta_{\rm ph}$).
The conventional FT is reproduced when $\beta_{\rm el}=\beta_{\rm ph}$.
Such a transport-without-bias behavior may also be generated by introducing asymmetric energy-dependent tunneling rates~\cite{sanchez2011a} or when a bosonic bath
couples directly to the QD occupation~\cite{entin_wohlman2010a}.
Also at finite bias voltages, the device may perform work by transporting electrons against an existing potential gradient, see Fig.~\ref{fig:Phon2in1}.
In our case, the required energy is provided by the temperature difference between the boson and fermion reservoirs.


\subsection{Bosonic transfer FT}

Evaluating the direct boson current between upper and lower baths requires to evaluate the full fluctuation theorem
at $\nU=+n_{\rm ph}$ and $\nD=-n_{\rm ph}$ -- disregarding the number of tunneled electrons at the other junctions.
Formally, this corresponds to
\bea
P_{n_{\rm ph}} &\to& \frac{1}{4\pi^2} \int\limits_{-\pi}^{+\pi} 
e^{\lambda_1(0,0,0,0,\chi_U,\chi_D) t -\ii n_{\rm ph} (\chi_U-\chi_D)} d^2\chi\,.\nn
\eea
We now consider the identities (again for similar electronic temperatures $\bl=\br=\beta_{\rm el}$ only)
\bea
\lambda_1(0,0,0,0,-\chi_U,-\chi_D) &&\stackrel{{\rm Eq.}~(\ref{eqn:symmetryI})}{=}\nn
\lambda_1\Big(
\ii \beta_{\rm el}(\eA-\ml),\ii\beta_{\rm el}(\eA+U-\ml),&&\nn
\ii \beta_{\rm el}(\eB-\mr),\ii\beta_{\rm el}(\eB+U-\mr),&&\nn
\chi_U+\ii\bu(\eB-\eA-\mU),&&\nn
\chi_D+\ii\bd(\eB-\eA-\mD)\Big)&&\stackrel{{\rm Eq.}~(\ref{eqn:symmetryII})}{=}\nn
\lambda_1\Big(
0,0,0,0,&&\nn
\chi_U+\ii\bu(\eB-\eA-\mU)-\ii\beta_{\rm el}(\eB-\eA+V),&&\nn
\chi_D+\ii\bd(\eB-\eA-\mD)-\ii\beta_{\rm el}(\eB-\eA+V)\Big)&&.
\eea
Finally, this implies that the bosonic transfer FT (see also e.g.~\cite{jarzynski2004a,saito2007b})
\bea\label{eqn:bosonic_transfer}
\frac{P_{+n_{\rm ph}}(t)}{P_{-n_{\rm ph}}(t)} &\to& e^{n_{\rm ph}\left[\bu (\eB-\eA-\mU)-\bd(\eB-\eA-\mD)\right]}
\eea
is not affected by the electronic transport at all, which is due to the fact that
for the boson-system coupling in our model, the coupling between transferred particles and energy is tight.


\subsection{Combined bosonic FT}

\begin{figure}[ht]
\includegraphics[width=0.48\textwidth,clip=true]{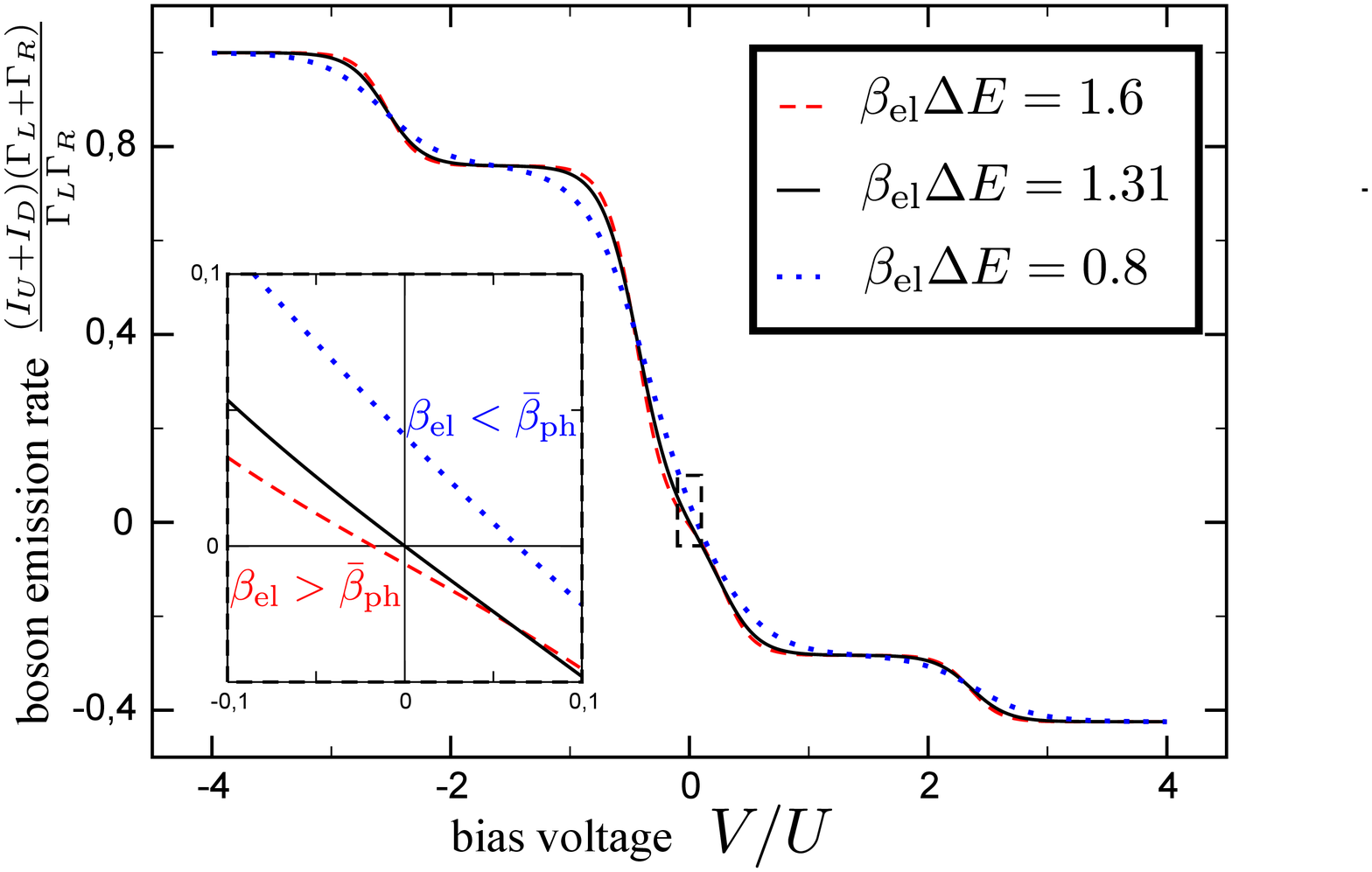}
\caption{\label{fig:Phonon_paper_neu}
(Color Online)
Renormalized phonon current from Eq.~(\ref{eqn:current}) versus bias voltage for different electron bath temperatures.
At zero bias voltage (compare zoomed inset), bosons may either be created (higher electronic temperatures, dotted blue curve) or 
absorbed (lower electronic temperatures, dashed red curve).
Parameters have been chosen as $\beta_U \Delta E \equiv \beta_U (\eB-\eA) =1.0$, $\beta_D \Delta E =2.0$
(leading to $\bar\beta_{\rm ph}\Delta E \approx 1.31$ at symmetric tunneling rates $\gu=\gd$), $\gl=\gr=\gu=\gd$, $\beta_D U=20$,
$\mu_U=\mu_D=0$ and $\mu_L=+V/2=-\mu_R$.
}
\end{figure}

When we do not differentiate in which of the bosonic baths phonons are counted 
$n_{\rm ph}=\nU+\nD$, the probability to count a given number of phonons is given by
\bea
P_{n_{\rm ph}} &\to& \frac{1}{2\pi} \int\limits_{-\pi}^{+\pi} 
e^{\lambda_1(0,0,0,0,\chi,\chi) t -\ii n_{\rm ph} \chi} d\chi\,.\nn
\eea
We now consider the identities (again for similar electronic temperatures $\bl=\br=\beta_{\rm el}$ only)
\bea
\lambda_1(0,0,0,0,-\chi,-\chi) &&\stackrel{{\rm Eq.}~(\ref{eqn:symmetryI})}{=}\nn
\lambda_1\Big(
\ii \beta_{\rm el}(\eA-\ml),\ii\beta_{\rm el}(\eA+U-\ml),&&\nn
\ii \beta_{\rm el}(\eB-\mr),\ii\beta_{\rm el}(\eB+U-\mr),&&\nn
\chi+\ii\bu(\eB-\eA-\mU),&&\nn
\chi+\ii\bd(\eB-\eA-\mD)\Big)&&\stackrel{{\rm Eq.}~(\ref{eqn:symmetryII})}{=}\nn
\lambda_1\Big(
0,0,0,0,&&\nn
\chi+\ii\bu(\eB-\eA-\mU)-\ii\beta_{\rm el}(\eB-\eA+V),&&\nn
\chi+\ii\bd(\eB-\eA-\mD)-\ii\beta_{\rm el}(\eB-\eA+V)\Big)&&\stackrel{{\rm Eq.}~(\ref{eqn:symmetryIII})}{=}\nn
\lambda_1\Big(
0,0,0,0,&&\nn
\chi+\ii\bar{\beta}_{\rm ph}(\eB-\eA)-\ii\beta_{\rm el}(\eB-\eA+V),&&\nn
\chi+\ii\bar{\beta}_{\rm ph}(\eB-\eA)-\ii\beta_{\rm el}(\eB-\eA+V)\Big)&&\,.
\eea
Finally, this implies that the combined bosonic FT
\bea\label{eqn:bosonic_combined}
\frac{P_{+n_{\rm ph}}(t)}{P_{-n_{\rm ph}}(t)} &\to& e^{n_{\rm ph}\left[\bar{\beta}_{\rm ph}(\eB-\eA)-\beta_{\rm el}(\eB-\eA+V)\right]}
\eea
is now affected by the electronic transport.
We have also calculated the total bosonic emission rate and find qualitative agreement with the FT, see Fig.~\ref{fig:Phonon_paper_neu}.

\subsection{Numerical Sanity Check}

We have also computed the fluctuation theorem in 
Eqns.~(\ref{eqn:electronic_transfer}),~(\ref{eqn:bosonic_transfer}), and~(\ref{eqn:bosonic_combined})
by performing the required one- or two-dimensional integration using the dominant eigenvalue numerically.
Within the boundaries of numerical accuracy, we have found complete agreement with our results (not shown).
Naturally, we have also tested the independence on electronic tunneling rates and the simpler dependence on an
average boson temperature.


\section{Summary}

We have investigated the multi-terminal fluctuation theorem for full counting statistics
for a four-terminal model including bosonic and fermionic channels and Coulomb-interaction as well as
phonon-assisted electron tunneling.
We find that under the Born-, Markov-, and secular approximations that under the assumption of a nondegenerate system spectrum
lead to the conventional rate equations, the complete FT is fully satisfied.
As these rate equations satisfy local detailed balance at each terminal separately, this
result was expected.

However, when not the complete information is gathered on all energy and matter fluxes, the
FT may be apparently violated, formally expressed e.g.\ by a renormalized bias voltage.
Thus, the modification of the FT may be qualitatively different from QPC back-action models but 
rather mimic the FT found for a Maxwell demon model.
In addition, when reservoirs at different thermal states couple identically to the system, they may 
act as a single bath at some average thermal state -- which may destroy universality of the FT (independence of
the tunneling rates).


\section{Acknowledgments}

We have profited from discussions with E. Schlottmann, R. Sanchez and G. Kie{\ss}lich and also gratefully
acknowledge financial support from the DFG (SCHA 1642/2-1).



\end{document}